# VCSEL-based CPO for Scale-Up in A.I. Datacenter – Status and Perspectives

M. Kohli and J. Teissier

*Abstract*[1]— The drastic increase of bandwidth demands in AI datacenters requires new solutions with low power consumption and high bandwidth densities. Further increasing the total bandwidth with copper interconnects is challenging, thus it is crucial to introduce optics into the scale-up network. For this purpose, the most important metrics are the energy efficiency of the total link in addition to the spatial bandwidth-density and reach. With the inefficiency of copper traces at high speeds, transitioning to co-packaged optics is no longer optional but essential. Here, we show that the mature VCSEL technology offers the ideal combination of low-cost, low-latency, high-reliability, and energy efficiency at all bitrates, thanks to their unique versatility and high wall-plug-efficiency.

## 1. INTRODUCTION

THE introduction of AI datacenters puts considerable pressure on data communication. In particular, all-to-all communication between GPUs within "scale-up" pods create an all-new market for short-reach. Demand for links with a length ranging from 0.3 to ~30 m has therefore skyrocketed and continues to increase as a function of the number of GPUs per rack and the size of the pod. This scale-up network represents more than 80% of the datacenter bandwidth in the training phase of the AI model. To further increase bandwidth-density, the only option with copper is to increase the bitrates at the cost of reach and power consumption. This requires increasing the speed of the multiplexing gearbox at the edge of the processor (SerDes). Furthermore, passive copper cables have a reach just below one meter at 200 Gbit/s PAM4. Increased reach for copper is only possible with active copper cables (ACC). However, ACCs typically consume more than their optical counterparts. A retiming is necessary, costing an extra 1.1 pJ/bit for three meters reach at 200 Gbit/s [1]. Furthermore, ACCs do not solve the bandwidth-density issue. The latest commercially available rack from Nvidia (NVL72) contains over 5000 passive copper cables for a total of 3.2 km with a cross section of 1.5 mm² [2], [3]. The limited space inside the rack ultimately defines the upper boundary of the bandwidth-density. Copper cables do not offer a future proof solution to answer the bandwidth increase inside AI datacenters. New solutions with a combination of high bandwidth, small footprint, and high energy efficiency are urgently needed.

*i. Current Copper Datacom Links*

Modern computers rely heavily on copper in CMOS processes as well as in signal traces on printed circuit boards (PCB). For current high-speed links, the data stream exits at the edge of the processor through hundreds of connections at slow bitrate (~4 Gbit/s) encoded in two levels (PAM2). To reduce the number of copper traces and the footprint on the PCB, a dedicated tile of the chip multiplexes these lanes to a faster signal rate (for example SerDes or UCIe). This is done at the cost of energy efficiency (EE), consuming approximately 0.25 pJ/bit [4]. For larger bandwidth per channel, the data can be encoded with higher modulation format, such as PAM4. This conversion costs an extra ~0.25 pJ/bit in EE at 200 Gbit/s. Furthermore, this tile typically adds digital signal processing (DSP) to compensate for the non-linearities in the copper traces. The amount of pre-distortion depends on the losses in the copper and can vary from an extra ~0.5 pJ/bit to reach the edge of the PCB to more than 4 pJ/bit for longer link [4]. Currently, a pluggable cable at the edge connects to a different board. On the receiver side, the process is mirrored to transfer the signal to the second processor.

Passive pluggable copper cables can transmit high-speed data signals without the need for extra electrical power. However, due to the longer effective length of the link, the DSP part of the SerDes adds a consumption of 4 pJ/bit. Maximum reach is limited to approximately one meter at 200 Gbit/s per lane for a total EE of 5 pJ/bit [4], [5].

*ii. The post-Copper Era*

In contrast to copper, optics can offer spectral, polarization and spatial multiplexing to optimize the trade-off between bitrate, energy consumption, and bandwidth-density. For the scale-up network, energy efficiency (EE) and latency are key. A move towards co-packaged optics (CPO), where the optical engine is co-packaged with the processor, lowers the electronic losses associated with the PCB. CPO also reduces the latency due to simpler electronics at the cost of requiring

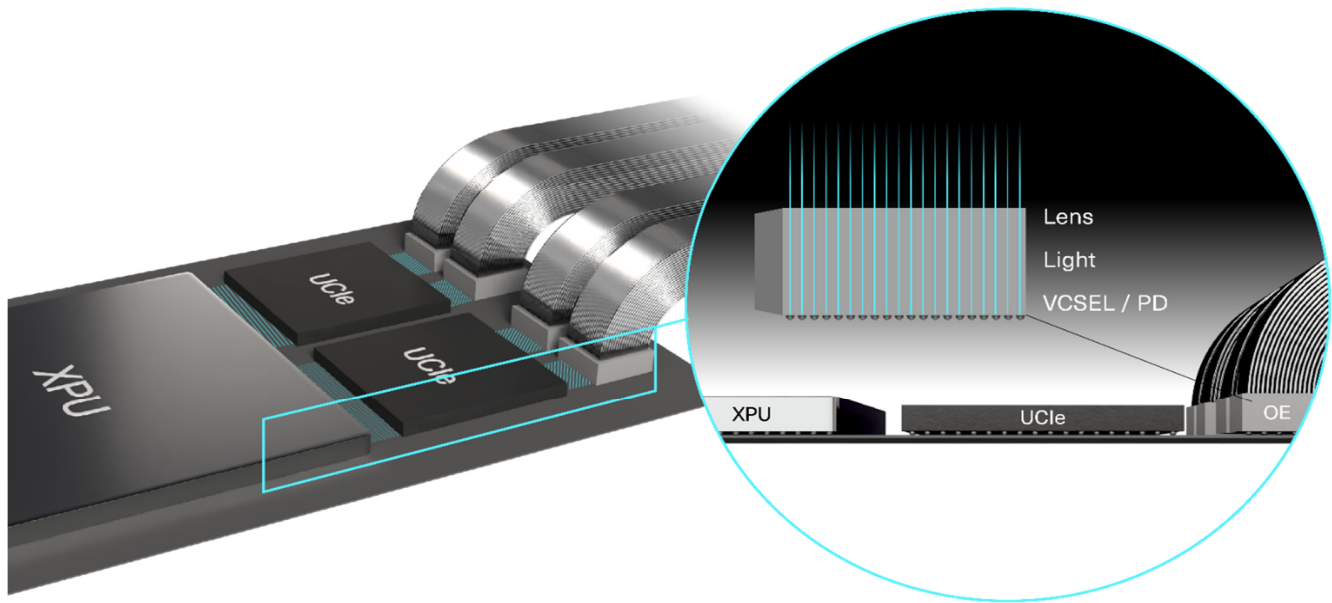

Figure 1. Conceptual representation of a UCIe-compatible VCSEL array link. The EIC tile on the interposer acts as the multiplexer in the electronic domain. Further integration is possible by combining the UCIe with the EIC of the optical engine.

higher optical bandwidth densities and excellent reliability at high temperatures.

Currently, the optical links can achieve the same speed per lane as copper in the "fast and narrow" (FaN) approach. However, continuously increasing the speed of optical links degrades the energy efficiency of the drivers and forces the use of digital signal processing (DSP). Alternatively, the bitrate could be slowed down while increasing the number of channels in a "slow and wide" (SaW) approach. This relaxes the complexity of the electronic chain. Lower speeds allow the removal of DSP and error correction electronics due to higher signal-to-noise ratio. This improvement in addition to the gain-bandwidth product of the receiver electronics significantly decreases the required optical power at the photodetector. Figure 1. Conceptual representation of a UCIe-compatible VCSEL array link. The EIC tile on the interposer acts as the multiplexer in the electronic domain. Further integration is possible by combining the UCIe with the EIC of the optical engine. visualizes the concept of a CPO WaS approach where the signal from the XPU is multiplexed by the UCIe tile before reaching the EIC+VCSEL/Photodiode tile. Multicore fibers are coupled to the VCSEL and Photodiode tiles.

The optimum operating speed per lane is a careful balance between several competing factors. First, there is a trade-off between simplifications due to lower speeds and static overheads such as turn-on voltages. Second, speed influences the balance between the reliability of electronics and that of the optical engine. A higher number of slower lanes tightens the reliability requirements on each individual lane, whereas reducing the per-lane power budget and simplifying the electronics improves for the overall failure rate. Finally, footprint and cost form an additional trade-off that must be optimized as part of the overall design. The bandwidth-density needs to be improved while keeping costs at an acceptable level. Here, we compare different solutions in the FaN (>100 Gbit/s) with 4 amplitude level (PAM4) encoding and WaS (<64 Gbit/s) in PAM2, e.g. non-return zero (NRZ), approaches. The results show that CPO VCSEL arrays are the main challenger for the scale-up network. VCSELs enable a unique combination of power efficiency, scalability, and reliability.

*iii. Competing Technologies*

Currently, we identify five main options to replace copper in CPO applications. These solutions, divided between the "*fast-and-narrow*" (FaN) and "*wide-and-slow*" (WaS) approaches, aim to deliver the aggregated bandwidth and energy efficiency required for the next generations of AI datacenters. Our focus lies on technologies that combine a compact footprint with high reliability, which are key enablers for scalable CPO integration:

1. Fast and narrow PAM4 VCSEL
2. Fast and narrow Silicon Photonics Microring (SiPho MRM) PAM4
3. Slow and wide NRZ SiPho MRM
4. Slow and wide NRZ microLED arrays
5. Slow and wide and VCSEL arrays.

The first approach is FaN PAM4 VCSELs, where CPO prototypes show 108 Gbps PAM4 [6]. To achieve 3.2 Tbit/s tiles, 32 VCSELs are required. Reliability of these active devices at high temperatures is going to be detailed in the following. VCSELs have been employed in pluggable for 25 years with hundreds of millions of units shipped, offering a very mature technology. Furthermore, achieving 200 Gbit/s VCSELs is in close reach [7]. The second approach is based on the FaN silicon photonics (SiPho) microring modulators (MRMs). Nvidia and Broadcom follow this option for the scale-out network. It relies on 200 Gbit/s/lane using the same SerDes as copper. The light source is an external pluggable laser, and the light is modulated by a silicon microring. Academic results show that the symbol rate of SiPho MRM can be pushed towards 400 Gbit/s/lane (160 GBd/s 6PAM) [8].

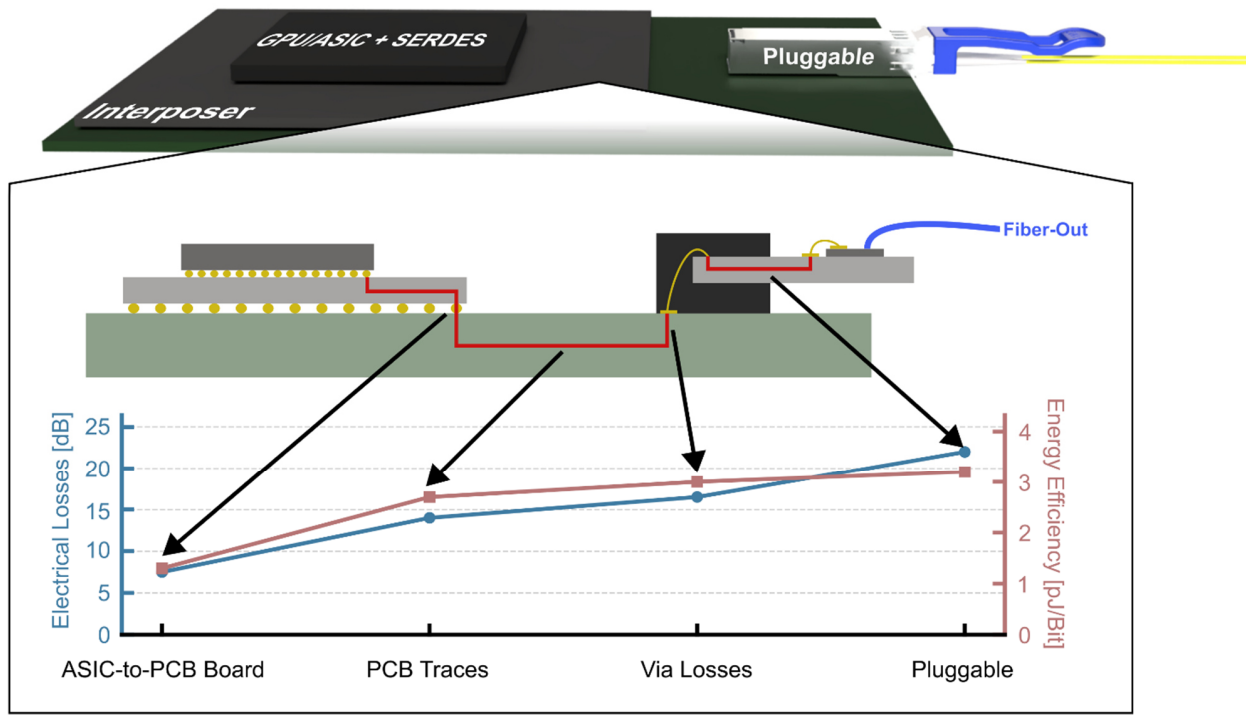

*Figure 2. Illustration of a pluggable datacom link at 200G per lane, with the cumulative electronic losses to reach the pluggable, correlated to the energy cost required to reach a pluggable. As an outcome, 22 dB propagation losses at 200G PAM4 to reach the pluggable is currently the state of the art. This translates into ~ 3.5 pJ/bit to just move the data from the processor to the pluggable.*

The advantage of this approach is the possibility of 3D co-integration with electronics, small dispersion in the O-band, and good reliability in comparison to SiPho pluggable (equal temperature, not at use condition) [9]. The challenge lies in increased costs due to the external laser pluggable, the lock-in of the microring resonance to the laser, polarization insensitive detection, and tight processing tolerance.

SiPho is also competitive with the WaS approach. Here, wavelength-multiplexing is implemented by adding many microrings with offset resonance frequency to the same bus waveguide. This approach benefits from very efficient detection in CMOS-compatible photodiodes. There are several start-ups working in this approach. For example Ayar Labs offers a UCIe compatible tile [10]. With this approach, there is an additional challenge of generating multiple laser wavelengths and the complex lock-in of the microring arrays to each color.

The fourth approach is microLED arrays, typically based on gallium nitride. The wavelength in the visible allows for efficient detection using standard CMOS. Other benefits include the small footprint of the LEDs and the well-established co-integration with CMOS. Microsoft and startups like Avicena are the main drivers behind the microLED. The coupling between microLEDs and fibers is challenging due to the broad spectrum and the wide emission profile. This slow-and-wide approach relies on spatially multiplexed communication inside imaging fiber bundles, typically used for image mapping, instead of wavelength multiplexing like SaW SiPho.

The fifth approach is slow-and-wide VCSEL arrays. Similarly to the microLED approach, VCSEL arrays are spatially multiplexed in a bundle or multicore fiber. In contrast to LEDs, lasers allow easier fiber coupling and enable extended reach. They also exhibit higher wall plug efficiency. The higher bandwidth of the VCSELs allows different data rates from 4 to >100 Gbit/s NRZ.

2. KEY TECHNOLOGICAL CONSIDERATIONS

*i. The Energy Efficiency Race*

Energy Efficiency (EE) is one of the most important aspects for scale-up networks. Reaching the edge of the board from the processor at 200 Gbit/s requires around 3.5 pJ/bit, including the SerDes and short-reach link DSP [4]. Figure 2 illustrates how each copper trace and each interface impacts the losses of the electrical signal and the EE it costs to correct for the induced signal distortions from the XPU to the pluggable. The reference pluggable link is the passive copper link, with a total EE of 5 pJ/bit due to extra DSP and a maximum reach of 1 m at 200 Gbit/s.

The dominant technology to reach 10 meters and beyond is the pluggable module with included DSP. They offer an EE around 10 pJ/bit in addition to the 3.5 pJ/bit for the SerDes at a total EE of 13.5 pJ/bit [5]. To increase the EE, the DSP in the module can be removed, in the optical engine is linear enough. This is done in the so-called linear drive pluggable optics (LPO) at the cost of more processing in the SerDes. Typical EE is around 5 pJ/bit in addition to the 5 pJ/bit in the SerDes (long-range protocol DSP) for a total EE of 10 pJ/bit. Another approach is to use WaS inside the pluggable. Microsoft for example demonstrated 3.6 pJ/bit at the pluggable with microLEDs for a total EE of 6.6-7 pJ/bit [11].

A logical next step to increase the EE is to remove losses in the copper to lower the power consumption at the SerDes level. CPO offers a way to cut most of the consumption at the board level, minimizing copper traces length and bump counts. Most mature solution for CPO is currently SiPho MRM. Currently, the EE of a SiPho optical engine from Nvidia and Broadcom is ~6.7 pJ/bit at 200 Gbit/s per lane [9]. Including the SerDes, the total EE is ~8 pJ/bit with room for improvement. Ideally, the DSP can be completely removed and replaced with analog pre-distortion if the entire link is linear enough. Such an approach has been demonstrated with VCSELs as a competing solution. At data rate of 108 Gbit/s, VCSEL CPO has shown an EE for the optical engine of 0.9 pJ/bit [6]. We expect the total EE of the tile to be around 2 pJ/bit.

Since optics allow for spatial and spectral multiplexing, further improvement in a WaS approach can be achieved by moving from PAM4 back to NRZ and decreasing the speed per lane. This improves the bit error rate (BER) due to lower SNR requirements and reduces complexity of the SerDes, saving around 0.5 pJ/bit. In this case, electronic losses are well optimized. This means that the EE is mainly dominated by the optical engine, breaking roughly even between the transmitter (Tx) and receiver (Rx). Main contender in SiPho NRZ is Ayar Labs, who offer UCIe connected tiles with a total EE <5 pJ/bit [10]. In the case of VCSELs, an EE of 1.54 pJ/bit is achieved at ~64 Gbit/s NRZ, with an electronic integrated circuit (EIC) designed for PAM4 [6].

VCSELs already demonstrated a bandwidth of ~15 GHz at a driving current as low as 0.5 mA . In combination with a voltage driven approach for the Tx and a very low capacitance photodiode to limit the energy consumption of the TIA on the Rx [12], we estimate that 32 Gbit/s VCSEL arrays can offer a total EE around 0.75 pJ/bit, taking benefit of UCIe, clock forwarding and other advantages offered by the SaW approach.

Finally, one can think of using a serializer-free approach. In academic environment, an impressive EE of 0.3 pJ/bit of the optical engine with SiPho MRM was demonstrated at 10 Gbit/s per lane [13]. Combined with the UCIe tile standard, it could potentially offer an EE around 0.5-0.6 pJ/bit. The challenge lies in the requirements for frequency combs with such a high number of wavelengths inside a laser pluggable.

With directly modulated emitters, however, there is no need for a laser pluggable at the cost of higher fiber core count. An example of this approach is done by Avicena with microLEDs in the visible wavelength. At a speed of 3.3 Gbit/s and a finely tuned Rx electronic circuit, they demonstrate 1 pJ/bit in back-to-back measurements of the optical engine. For such finely tuned systems, the wall plug efficiency of the emitters is crucial. In the case of microLEDs, the efficient Rx electronics and threshold-less operation cannot fully compensate for the low emitter's wall plug efficiency of 1.2% [14]. Furthermore, GaN LEDs also requires a drive voltage around 4.4 V. Furthermore, the large fiber attenuation requires higher output power than in back-to-back experiments. Thus, microLEDs achieve 3.6 pJ/bit in the optical engine at limited bitrates of 1.6 Gbit/s [11]. VCSELs, on the other hand, do need to operate much above threshold current However, by removing the TIA and employing resistive loading of the photodiode [15], it should be possible to achieve 0.5 pJ/bit for VCSEL-based links at 4 Gbit/s.

Figure 3 summarize the estimated EE for short-reach communications within AI datacenters, based on the numbers from this chapter. VCSELs offer a unique combination of bandwidth and wall plug efficiency (>20 %) over the entire bitrate range. The 3-dB bandwidth can exceed 15 GHz at 500 $\mu$A [12]. At higher bitrate more optical power is required at the receiver. However, this extra power is compensated for by the bitrate. For the WaS approach, optimization toward lower thresholds can help to achieve the optimal EE of the optical engine below 1 pJ/bit. In contrast to SiPho MRM, there are no requirements for lock-in electronics and frequency combs, thus VCSELs provide the least complex solution to replace copper.

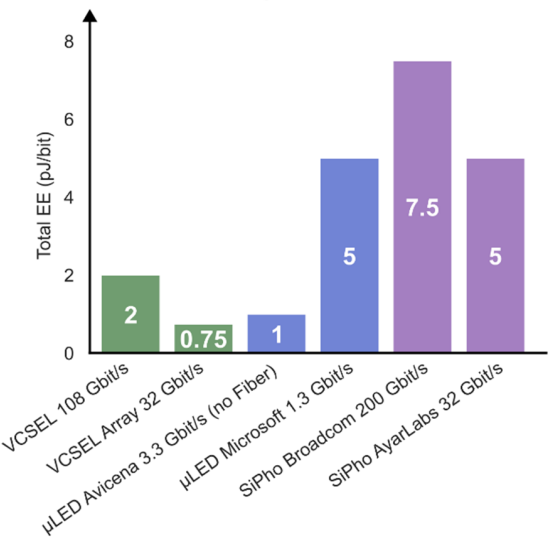

Figure 3. The estimated energy consumption of the main technologies for short-reach communications within AI datacenters

*ii. Footprint*

Historically, the main footprint concern was placed on the beachfront density of the GPU, which is defined as the bandwidth-density exiting the edge of the chip. Recent development in vertical coupling, however, moves the focus towards bandwidth-density in Tbit/s/mm$^2$. All approaches highlighted here can fulfill the current CPO bandwidth-density requirements of >0.5 Tbit/s/mm$^2$. The best bandwidth-density can be achieved with 3D co-integration, where the optical engine and the EIC are stacked on top of each other [16]. The versatility of EIC in terms of technology, node and layer stack relieves the potential issue of the bandwidth-density from the electronics side. With lower speed, the EIC's complexity decreases and counteracts the challenge of increased number of lanes. For the optical part of the Tx and Rx, current state-of-the-art values in SiPho MRM range from 0.5 to ~1 Tbit/s/mm$^2$, excluding the SerDes footprint. For microLEDs the bandwidth-density is mostly limited by the pitch and the speed-per-lane. At current values of 50 µm and 4 Gbit/s, the bandwidth-density is at 0.8 Tbit/s/mm$^2$ assuming similar footprint for Tx and Rx. In the case of VCSELs, similar pitches are achievable. Together with higher speed, this results in much higher bandwidth-density. For example, at a moderate speed of 32 Gbit/s and a 50 µm pitch VCSEL arrays achieve 6.4 Tbit/s/mm$^2$.

*iii. Reach*

*Required reach for scale-up*
There is a trade-off between the size of the pod for increased computer power and the time-of-flight due to the distance in the communication links. Typically, pod sizes are between 1 to 30 m. In the case of current Nvidia rack type GB200 NVL72, the scale-up network is located within the rack connecting 72 GPUs in an all-to-all connection. These links are currently well below 2 m due to the reach of passive copper cables at 200 G/lane [17]. In the case of Huawei, 384 GPUs are in an all-to-all connection for the scale up [18] and the size continuously increase, with Huawei and Google each planning more than 9000 GPU in all-to-all interconnect "SuperPods" [19], [20].

*Reach Limitations*
Reach limitations in fibers is a convolution between the limits caused by absorption, modal dispersion, and chromatic dispersion. Each technology has a specific balance between these three limitations. SiPho with the 1310 nm single-mode fibers allow for longest reach.

The absorption in single-mode fibers used in SiPho is below 0.3 dB/km and is therefore negligible here. Similarly, the losses in a typical multimode fiber (e.g. OM4) for the 850 to 1060 nm range are low for scale-up (2.3 dB/km). The fiber bundles used in WaS for VCSELs and microLED arrays consist of hundreds of cores. The plastic bundle fibers used for microLED arrays show typical transmission values around 3 dB for 5 m at a wavelength of 420 nm. For VCSEL arrays, however, one can employ multicore fibers with silica cores, which offer similar losses as the OMx fiber type. New fiber types may lower absorption UV, but Rayleigh scattering losses limit propagation losses at short wavelengths.

The multimode fibers (OMx), used traditionally for VCSELs, have different propagation speeds for each mode. This modal dispersion limits the range to around 150 m for the 100 Gbit/s 850 nm VCELs in FaN. The WaS approach suffers less from this limitation. Custom fiber bundles of single-mode cores with very high-density pitch will virtually cancel this dispersion [12]. Chromatic dispersion is a similar concept, every wavelength having a different propagation speed. The narrow linewidth of the DFB laser allows SiPho to reach multiple kilometers. VCSELs have intrinsically a wider linewidth due to the direct modulation. This limits the reach to ~1-2 km for FaN VCSELs. The broad nature of the microLED spectrum, with a spectral width above 20 nm [14], together with the strong achromaticity of the fibers at 420 nm, limits the bitrate to 1.6 Gbit/s for 30 m to keep the BER <10$^{-4}$ [11].

Figure 4 illustrates the eye diagrams of a good channel in (a), a lossy channel in (b) and a distorted channel due to bandwidth limitations or dispersion in (c).

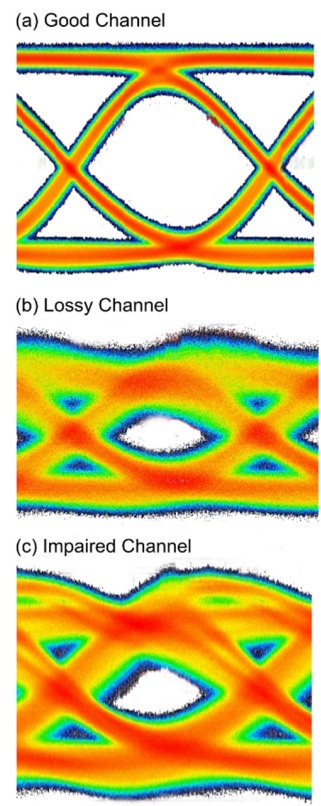

Figure 4. Examples of eye diagrams with different channel restrictions. (a) Eye diagram after an ideal optical channel. (b) Eye diagram with a lossy channel, e.g. fiber loss or electrical loss. (c) Eye diagram after a severely impaired channel, e.g. dispersion or bandwidth limitations.

*iv. Reliability*

Training phase of a model is the most sensitive phase for the network. Every error requires a restart of the training phase for the entire scale-up network from the last backup point. With a training period of about a month and costs above 100 M$, any improvement in the meantime between failure saves millions of dollars [9]. For the optical engine, we divide this chapter between the non-definitive failures, that do not

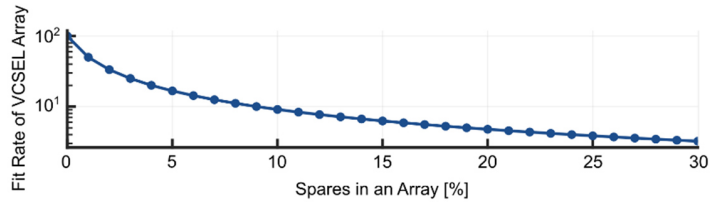
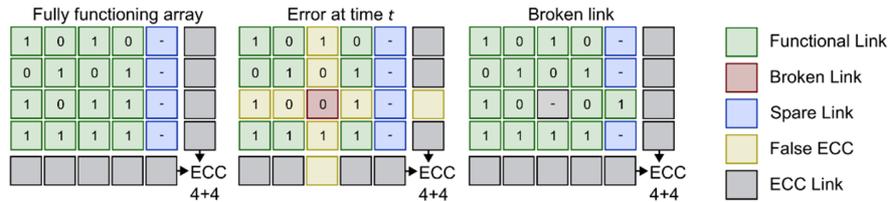

Figure 5. (a) Calculated FIT rate of a VCSEL array as a function of percentage of spares. A total FIT of 100 is assumed for the array without spares. (b) Visual representation of a simple forward error correction code with hot swaps. This scheme allows for zero link flaps as demonstrated by Microsoft. More complex Error Code Corrections are implemented in HBM.

require replacement of the optical engine, and the device failure, where the optical engine needs to be exchanged or serviced.

*Bit Error Rate and Link-Flaps*
Interruptions in the link can come from a glitch in the data stream, causing an error in the readout. The used metric in this case is the bit error rate (BER) and is expected to be below $10^{-12}$ for an "error-free" datacom link in NRZ, or below $10^{-6}$ in PAM4. The interruptions can also arise from changes in the optical power sent or received, called a link-flap. In pluggable, they can be caused by dust particles passing through the free-space optical path. Meta and Broadcom recently demonstrated in CPO more than 1 million device hours (0.8 Gbit/s tile) without a single link-flap in stressed-environment, while maintaining a BER $<10^{-7}$ in PAM4 [9]. In the WaS approach, these interruptions can be reduced to virtually zero as demonstrated by Microsoft [11]. This is achieved by adding a simple forward error correction using dedicated extra channels for it.

*Device Failure*
Moving from pluggable to CPO architecture imposes stringent reliability requirements, since the optical engine becomes non-replaceable. Ideally, the optical engine shows failure in time (FIT = fails per $10^9$ hours) rates significantly below the electronics. Meta shows around 5 times improvement in mean time between link failures (MTBF) for CPO in comparison to pluggable. The Bailly 51.2 Tbit/s CPO switches from Broadcom with SiPho MRM demonstrated below 66 FIT for a 0.8 Tbit/s CPO tile for unserviceable fails [9]. This improvement together with the reduction of link-flap allows for fewer restarts and thus decreases the cost of training by ~50%. In the case of microLEDs, a typical FIT of 0.1 per emitter is expected [11]. At a speed of 2 Gbit/s, this results in a total FIT of 160 for a 3.2 Tbit/s tile for the emitters only. The advantage of WaS approaches, especially for microLED and VCSEL arrays, lies in the easy implementation of spares. With 15 % of spares, the total link is expected to drop below 10 FIT rate. Coherent's 100G VCSELs show a better FIT rate of approximately 0.03 FIT for single emitters at 120°C and 4 mA drive current and 0.07 FIT at 85°C and 5 mA current. Assuming that the directly modulated laser accounts for ~40% of the fails [21], we take 0.2 FIT as an upper boundary for the entire link. It corresponds to an aggregated FIT of 6.4 with 32x100G links (3.2 Tbit/s tile). Adding only two spares would give around 1 FIT. In the case of slow and wide, assuming 512 links, the worst case aggregated FIT rate would be ~100 FIT. Adding 10% of spares reduces the FIT to approximately 10. Figure 5(a) shows the effect of increasing the percentage of spares with a starting overall FIT of 100. Optimizing the VCELS for WaS (slower speed, lower threshold current, lower output power and thus lower drive currents) will only improve reliability. Furthermore, Microsoft shows that in WaS [11], the combination of spares and error code correction allows not only to recover from a short link's interruption but also keep the link active on the long term by hot swapping any defective link.

Microsoft approach predicts more resilient links than copper. We believe that similarly, this approach can be implemented for VCSEL arrays. Figure 5(b) shows the visual representation of a error-correction code in a 2D array with spares and hot-swaps.

*E. Cost*
To compare the costs between the different approaches, the full link must be considered, including the SerDes/UCIe express tile. First, the cost of the optical engine can be reduced moving from fully retimed pluggable to LPO. The reason for this reduction is the cost of the DSP in the pluggable, which represents the single most expensive device on the assembly. Costs for 800G pluggable go from 3.5 $/Gbps down to ~0.05 $/Gbps forecasted in volumes for LPO. Including the

Table 1: Overview of different optical communication technologies for AI datacenters

| General Parameters | VCSEL 108 Gb/s PAM4 | VCSEL 32 Gb/s Array NRZ | μLED[(2)] [14] 3.3 Gb/s Array NRZ | μLED [11] 1.3 Gb/s Array NRZ | SiPho 200 Gb/s PAM4 Broadcom [24] | SiPho 32 Gb/s NRZ AyarLabs [10] |
|---|---|---|---|---|---|---|
| Reach (m) | 100 | 100 [22] | 0 [22] | 20 | km | km |
| Power for 3.2T total link (W) | <6.2 [6] | <2.4 [6]* | 3.2 | <16 | <24 | <16 |
| Optical Engine EE (pJ/bit) | 0.9 [6] | < 0.5 [6]* | ~1 | 3.6 | 5.4 | ~ 3.5 |
| # emitters for 3.2T | 32 | 100 | ~1000 | ~2500 | 16 | 200 |
| Max Operating Temp (°C) | >120 | >120 | >125 | >125 | N.C. | >85 |
| Pre-FEC Bit Error Ratio | <$10^{-6}$ [6] | <$10^{-12}$ | <$10^{-12}$ | <$10^{-12}$ | $10^{-7}$ [25] | <$10^{-12}$ |
| Latency | Medium (FFE) [6] | Low | Low | Low | Medium (FEC) | Low |
| Relative Cost per bit | Low/No DSP [6] | Low/No DSP | Low, integrated PD | Low | Medium | Medium |

*based on expectation of a direct drive without equalization

SerDes/UCIe tile significantly increases the cost. Second, the move towards CPO allows further cost savings thanks to casing removal and co-integration of the driver and TIA. In the case of WaS simpler gearbox further lowers the cost. For the SiPho MRM, the cost remains higher due to the laser pluggable (ELSFP) in comparison to directly driven methods (VCSELs, LEDs). Similarly to power consumption, the lowest cost per Gbit/s should be achievable with directly driven emitters at the speed of the CPU and GPU if the custom-fibers' cost goes down significantly. The optimal trade-off will depend on the average link length. Expecting costs as low as copper for the high volumes required for scale-up is possible if the cost of the SerDes/UCIe tile is factored in.

### 3. Conclusion

We summarize in Table 1 the key metrics of the approaches studied above. We focus on BER<$10^{-12}$ for NRZ (Error-Free Datacom threshold) and BER<$10^{-6}$ for PAM4, to minimize the electronic equalization that increases power consumption and latency. The last line considers the total link EE from processor to processor in the case of a scale-up network. We emphasize that the sub pJ/bit for microLED was only demonstrated in back-to-back experiments to the best of our knowledge, and that adding fiber-arrays significantly degrades the performance, as highlighted by Microsoft [10].

VCSEL based solutions offer state-of-the-art energy efficiency at all bitrates, thanks to their unique versatility and wall-plug-efficiency. They provide a low-cost, low-latency and high-reliability solution that can follow each requirement of a scale-up deployment. Since VCSELs are directly modulated light sources, they enable parallel optical architecture where a single channel failure remains isolated and does not propagate to others. In contrast, SiPho implementations often place multiple resonant rings on a single waveguide, where the failure of one element can compromise adjacent channels. This makes VCSELs currently the only solution to offer true redundancy across all bitrates while achieving the reach and energy efficiency targets. VCSELs can be configured for a "fast-and-narrow" approach—supporting 100 Gbit/s or 200 Gbit/s PAM4 links—or optimized for a "slow-and-wide" architecture using custom multicore fibers. To date, VCSELs are the only technology to demonstrate a production-ready CPO solution capable of operating below 1 pJ/bit for a scale-up fiber link. While "fast-and-narrow" links allow for copper interoperability with better key metrics, we show that a "slow-and-wide" VCSEL arrays further enhance key metrics such as energy efficiency and mean time between link failures, benefiting from inherent redundancy and advanced coding techniques.

### 4. Acknowledgment

The authors would like to express their sincere gratitude to Wilfried Maineult, Stefano Tirelli, Evgeny Zibik, Simon Hoenl, Julien Boucart, Mirko Hoser, Michael Moser, and Michele Agresti for their valuable technical discussions, critical feedback, and careful review of the manuscript, which significantly contributed to the quality of this work. The authors further acknowledge Evgeny, Zibik, Michael Moser, Michele Agresti, Karlheinz Gulden, Vipul Bhatt, and Giovanni Barbarossa for their guidance, expertise, and continued support throughout the development and execution of this research. The authors also thank Brian Lee for providing selected visualizations used in this work.